\documentclass[preprint,preprintnumbers,amsmath,amssymb]{revtex4}
\usepackage[abs]{overpic}
\usepackage{graphicx,subfigure}
\preprint{HUPD1307}
\begin{document}
\def\nn{\nonumber}
\def\beq{\begin{equation}}
\def\eeq{\end{equation}}
\def\bei{\begin{itemize}}
\def\eei{\end{itemize}}
\def\bea{\begin{eqnarray}}
\def\eea{\end{eqnarray}}
\def\ynu{y_{\nu}}
\def\nub{{\bar{\nu}}}
\def\Hp{{H^+}}
\def\ep{{e^+}}
\def\em{{e^-}}
\def\ydu{y_{\triangle}}
\def\ynut{{y_{\nu}}^T}
\def\ynuv{y_{\nu}\frac{v}{\sqrt{2}}}
\def\ynuvt{{\ynut}\frac{v}{\sqrt{2}}}
\def\s{\partial \hspace{-.47em}/}
\def\ad{\overleftrightarrow{\partial}}
\def\ss{s \hspace{-.47em}/}
\def\pp{p \hspace{-.47em}/}
\def\pdx{\frac{\partial}{\partial x}}
\def\bos{\boldsymbol}
\title{
Time variation of particle  and anti-particle asymmetry\\
in an expanding universe}
\author{Ryuichi Hotta, Takuya  Morozumi}
\address{Graduate School of Science, Hiroshima University
Higashi-Hiroshima, 739-8526, Japan}
\author{Hiroyuki Takata}
\address{Tomsk State Pedagogical University, Tomsk, 634061, Russia}
\begin{abstract}
Particle number violating interactions wash out the primordial 
asymmetry of particle number density generated by some interaction 
satisfying Sakharov conditions for baryogenesis.
In this paper, we study how the primordial
asymmetry evolves in time under the presence of particle number 
violating interactions and in the environment of expanding universe.
We introduce a complex scalar model with particle number violating mass terms 
and calculate the time evolution of the particle number density 
with non-equilibrium quantum field theory.
We show how the time evolution of the number density depends on
parameters, including the chemical potential related with the particle number, 
temperature, 
size of the particle number violating mass terms, and 
the expansion rate of the universe.
Depending upon whether
the chemical potential is larger or smaller than the rest mass of the scalar
particle,
behaviors of the number density are very different to each other.
When the chemical potential is smaller than the mass,
the interference among the contribution of oscillators with
various momenta reduces
the number density in addition to the dilution due to the expansion of universe.
In opposite case, the oscillation of the particle number density
lasts for a long time and the cancellation
due to the interference does not occur. 
\end{abstract}
\maketitle
\section{introduction}
Exploring the origin of the matter and anti-matter asymmetry of our universe, 
its production mechanism and time evolution are very important issues.
In many scenarios of baryogenesis \cite{Sakharov:1967dj} \cite{Yoshimura:1978ex} and leptogenesis \cite{Fukugita:1986hr},
baryon number $(B)$ and lepton number $(L)$ 
interactions are required so that 
the primordial asymmetry of the particle number can be generated.  
After it is generated, the
particle number violating interactions must be frozen. Otherwise,
the primordial asymmetry created will be washed out. In this regards,
in the context of leptogenesis \cite{Fukugita:1986hr},
there are studies of the effect
$\Delta L=2$ operator of the mass dimension $5$ 
on the  primordial $\frac{B}{3}-L_i$ ($i=e, \mu, \tau$) asymmetries.
If the coefficient of the operator is too large, the primordial
asymmetries will be completely washed out. 
Since the same operator generates the Majorana mass matrix for 
light neutrinos at low energy, 
constraints on its elements are obtained from the
condition that a leptogenesis scenario succeeds. One can also
argue whether they
are compatible  with neutrino masses, lepton mixing matrix, and 
the experimental limit on the neutrino-less double beta decay rate
\cite{Hasegawa:2003vh}.

We introduce a scalar model with particle number violating mass terms
to investigate 
the time dependence of the particle number density in expanding universe,
where the scale factor has arbitrary time dependence.
In numerical study,  
we focus on the case that the scale factor grows exponentially 
and study how a given initial particle number asymmetry evolves under the influence of the particle number 
violating mass terms and the expansion. 
The scalar field is written in terms of a complex Klein Gordon field 
and one can 
identify the time component of U(1) 
current as a particle number density.
Baryogenesis with a complex field has been discussed in several 
literatures \cite{Dimopoulos:1978kv},
\cite{Affleck:1984fy}, 
\cite{Takeuchi:2010tm},
where the time derivative of the phase of the scalar 
is identified with the baryon number density.

We adopt the non-equilibrium field theory which has been developed in
the literatures
\cite{Schwinger:1960qe},\cite{Bakshi:1962dv}, 
\cite{Bakshi:1963bn},\cite{Keldysh:1964ud},\cite{Ramsey:1997qc},
\cite{Calzetta:1986cq}
so that one can study the time evolution of the expectation value of
the particle number density.  
We employ functional method since one can naturally extend the present study so that interactions
and condensates are incorporated.   The present work, therefore,
serves as a starting point when we include interactions besides 
the terms quadratic with respect to the field. 
Despite of no interactions beyond quadratic terms, under the environment
of the expanding universe,
the
initial condition with non-zero asymmetry and 
the particle number violating mass terms 
lead to non-trivial time evolution of
the particle and anti-particle asymmetry.
In the expectation value, the weight of each state is specified by a
density matrix. 
The density matrix is written with the grand canonical form and
it is specified with temperature and chemical potential.
The functional form for the density matrix with non-zero chemical potential
is constructed explicitly. 
In our study, the primordial asymmetry of the particle number density
is given by choosing the value and the sign of the chemical
potential.

We derive formulae for the expectation value of the 
particle number density in an analytic form.  For exponentially expanding 
universe,
the formulae are written 
with Hankel functions. 
The various limiting cases, e.g., the case of the vanishing and/or small
expansion rate and 
the case for vanishing particle number violating mass, etc., can be
easily obtained.
In numerical study, one can change the coefficient of the particle number violating mass term and
the expansion rate of the universe. One can also change initial conditions 
by specifying the temperature and the chemical potential in the density matrix. 
Therefore, in an unified way, 
one can study its time evolutions for the cases with 
different sets of parameters.

The paper is organized as follows.
In section II, Lagrangian for the scalar model is given. The initial
density matrix is also specified. In section III, 
using a two particle irreducible effective action, we solve the Schwinger Dyson equation for Green functions and obtain the
particle number density at arbitrary time. In section IV, we present the
numerical results and section V is devoted to summary.
In appendix A , 
the derivation of particle number
density for small expansion rate is given and in appendix B, that for the
vanishing limit of the particle number violating mass term is obtained.    
\section{The complex scalar model with U(1) breaking}
We start with a complex scalar model including a soft U(1) symmetry
breaking mass term. The time component of the U(1) current is a particle number density,
\bea
S&=&\int d^4 x \sqrt{-g} {\cal L},\nn \\
{\cal L}&=&g^{\mu \nu} \nabla_\mu \phi^\ast \nabla_\nu \phi
+\frac{B^2}{2} (\phi^2+\phi^{\ast 2}) \nn \\
&-&m_{\phi}^2 |\phi|^2+(\frac{\alpha_2}{2} \phi^2 + h.c.) R
+\alpha_3 |\phi|^2 R.
\eea
U(1) breaking terms are denoted with their coefficients
; $B$ and $\alpha_2$.  The metric
$g_{\mu \nu}$ is given by that of Friedmann Robertson Walker,
\bea
g_{\mu \nu}=(1,-a(t)^2,-a(t)^2,-a(t)^2).
\eea
The Riemann curvature is given as $R=12 H^2$ with
$H=\frac{\dot{a}}{a}$.
When
$\alpha_2$ and $B$ are real parameters, the mass eigenstates
of the scalar 
are the real part and the imaginary part of the complex scalar $\phi$.
By decomposing it into a real part $\phi_1$ and an imaginary part $\phi_2$
as
$\phi=\frac{\phi_1+i \phi_2}{\sqrt{2}}$, their masses
are given as follows,
\bea
\tilde{m_1}^2(x^0)&=&m_\phi^2-B^2 - 12 \alpha_3 H^2- 12 \alpha_2 H^2, \nn \\
\tilde{m_2}^2(x^0)&=&m_\phi^2+B^2 - 12 \alpha_3 H^2 + 12 \alpha_2 H^2.
\label{eq:mtilde}
\eea
The current associated with U(1) transformation 
$\phi^\prime=\phi e^{i \theta}$ is \cite{Dimopoulos:1978kv},
\bea
j_\mu&=&i (\phi^\dagger \partial_\mu \phi-\partial_\mu \phi^\dagger \phi)
,\nn \\
&=&\phi_1 \partial_\mu \phi_2-\phi_1 \partial_\mu \phi_2. 
\label{eq:current}
\eea
Next we study the density matrix which specifies the initial state.
Since we have non-vanishing primordial asymmetry of the particle number 
density, the statistical density matrix has the following form with non-zero
chemical potential, 
\bea
\rho=\frac{e^{-\beta(H_0-\mu N)}}{{\rm tr}e^{-\beta(H_0-\mu N)}},
\label{eq:densitymatrix}
\eea
where $H_0$ corresponds to the Hamiltonian obtained by taking the 
U(1) breaking terms
and curvature dependent terms turned off and $N$ is a particle number operator
defined as follows,
\bea
N=\int d^3 x \sqrt{-g} j^0.
\label{eq:N} 
\eea
The expectation value of the U(1) current is written with the
density matrix in Eq.(\ref{eq:densitymatrix}).
\bea
\langle j_\mu(X) \rangle ={\rm tr}(j_\mu(X) \rho).
\label{eq:expectj}
\eea
In section III, we compute the expectation value with
the Green function of 2 PI (particle irreducible) formalism. 
From the definition of 
the U(1) current in Eq.(\ref{eq:current}), 
the expectation value defined in Eq.(\ref{eq:expectj}) 
can be written in terms of 
the Green function, 
\bea
G_{12}^{12}(x,y)\equiv{\rm tr}(\phi_2(y) \phi_1(x) \rho).
\label{eq:green}
\eea
The resulting formulae for the expectation value of the
current is given as follows,
\bea
\langle j_\mu(X) \rangle =(\frac{\partial}{\partial x^\mu}-\frac{\partial}{\partial y^\mu})
G_{12}^{12}(x,y) \Bigr{|}_{x=y=X}.
\eea
\section{Schwinger Dyson equation from 2 particle irreducible effective action}
In this section, we derive 2 PI effective action
and obtain the Schwinger Dyson equations for Green functions.
By solving the Schwinger Dyson equations, we obtain
an analytic form for the expectation value of
the current for the case that the scale factor of the universe 
of arbitrary time dependence.

2 PI effective action in curved space time for O(N) theory is
derived in \cite{Ramsey:1997qc} and the method employed 
can be also applied
to the present model.   
In 2 PI formalism, one first introduces non-local source term denoted by $K$ in addition to the usual local source term $J$.
\bea
&& e^{i W[J,K]}=\nn \\
&& \int d \phi e^{i[S+ i \int \sqrt{-g(x)} d^4 x
c^{ab}J_i^a \phi_i^b+ \frac{1}{2} 
\int d^4 x d^4 y \sqrt{-g(x)} c^{ab} c^{cd} \phi_i^b(x) 
K^{ac}_{ij}(x,y) 
\phi_j^d(y) \sqrt{-g(y)}]}.
\eea
where $c^{ab}$ is the metric of in-in formalism \cite{Calzetta:1986cq} 
and $c^{11}=-c^{22}=1$
and $c^{12}=c^{21}=0$.

The Legendre transformation of $W$ leads to the 2 PI effective action,
which is a functional of Green function.
\bea
\Gamma[G,\hat{\phi},g]=S[\hat{\phi},g]+\frac{i}{2} {\rm Tr} {\rm Ln} G^{-1}+
\frac{i}{2} \int d^4 x \int d^4 y M^{ab}_{ij}(x,y) G^{ab}_{ij}(y,x),
\eea
where $S$ and $M^{ab}_{ij}$ are given by, 
\bea
S[\hat{\phi},g]&=&\frac{1}{2} \int d^4 x
\sqrt{-g(x)} 
c^{ab} (g^{\mu \nu} \nabla_\mu \hat{\phi}^a_i 
\nabla_\nu \phi^b_i-\tilde{m}_i^2 
\hat{\phi}^a_i \hat{\phi}^b_i)
,\\
i M^{ab}_{ij}(x,y)&=&-c^{ab} \delta_{ij}
\sqrt{-g(x)} (\nabla_x^\mu \nabla_{\mu}^x+ \tilde{m}_j^2) 
\delta^4(x-y).
\eea
The variation of the 2 PI effective action with respect to
the scalar field $\hat{\phi}$ leads to,
\bea
\frac{\delta \Gamma}{\delta \hat{\phi}^{a}_{i}(x)}=
-\sqrt{-g(x)}c^{ab}\{J_i^b(x)+
c^{cd} \int d^4 z \sqrt{-g(z)}
K^{bc}_{il}(x,z) \hat{\phi}^d_l(z)\},  
\eea
and one obtains the following equation of motion for the scalar field
$\hat{\phi}_i$.
\bea
c^{ab} (g^{\mu \nu} 
\nabla_\mu \nabla_\nu +\tilde{m}_i^2)\hat{\phi}^b_i
= c^{ab}\{J_i^b(x)+
c^{cd} \int d^4 z \sqrt{-g(z)}
K^{bc}_{il}(x,z) \hat{\phi}^d_l(z)\}.
\eea
When the single source term $J$ vanishes, 
the equation of motion for
$\hat{\phi}$ is homogeneous and linear with respect to $\hat{\phi}$.
Therefore $\hat{\phi}=0$ is 
a solution in this case.
The variation of the 2 PI effective action with respect to Green function
$G$ is the source term $K$,
\bea
\frac{\delta \Gamma}{\delta G^{ab}_{ij}(x,y)}=-\frac{1}{2} c^{ac} c^{bd} \sqrt{-g(x)}K^{cd}(x,y)_{ij} \sqrt{-g(y)}.
\label{eq:dg}
\eea
Eq.(\ref{eq:dg}) leads to two differential equations,
\bea
(\nabla_x^\mu \nabla_{\mu}^x+ \tilde{m}_m^2) G_{mn}^{a b}(x,y)&=&
-i\frac{1}{\sqrt{-g(x)}}c^{a b}\delta_{mn} \delta(x-y) \nn \\
&+& \int d^4 z 
K^{a c}_{ml}(x,z) \sqrt{-g(z)} c^{c d} G_{ln}^{d b}
(z,y),\nn \\
(\nabla_y^\mu \nabla_{\mu}^y+\tilde{m}_n^2) G_{mn}^{a b}(x,y)&=&
-i c^{a b}\delta_{mn} \delta(x-y)\frac{1}{\sqrt{-g(y)}} \nn \\
&+& \int d^4 z G_{ml}^{a c}(x,z) c^{c d} \sqrt{-g(z)} K_{ln}^{d b}(z,y).
\label{eq:twodif}
\eea
The non-local source term $K$ is related to
the functional representation of the initial density matrix $\rho$
introduced in Eq.(\ref{eq:densitymatrix}) \cite{Calzetta:1986cq},
\bea
\langle {\phi}^{1}|\rho|{\phi}^{2} \rangle
= C \exp[\frac{i}{2} \int \int d^4 x d^4 y \sqrt{-g(x)}   
c^{ab} \phi_i^b(x) K_{ij}^{ac}(x,y) c^{cd} \phi_j^{d}(y) \sqrt{-g(y)}],
\eea
where $C$ is a normalization factor and is determined 
so that the density matrix is normalized as ${\rm tr} \rho=1$.
$K$ is non-zero only if both $x^0$ and $y^0$ are the initial time.
The resulting $K$ has the following form,
\bea
K_{ij}^{ab}(x,y)=-i
\delta(x_0) \delta(y_0) \kappa_{ij}^{ab}({\bf x-y}),
\eea
where $\kappa$ specifies the space dependent part. Since it is 
invariant under translation, one can carry out the Fourier transformation on it.
\bea
\kappa({\bf x})=\int \frac{d^3 k}{(2 \pi)^3}
\kappa({\bf k})e^{-i {\bf{k \cdot x}}}.
\label{eq:kappa}
\eea
Let us derive the functional representation for the density matrix
of Eq.(\ref{eq:densitymatrix}) and determine $\kappa$.
\bea
&& \langle {\phi}^{1}|\exp(-\beta (H_0-\mu N))|{\phi}^{2} \rangle
=\exp(\beta \mu \hat{N}) \langle {\phi}^{1}|\exp(-\beta H_0)|{\phi}^{2} 
\rangle
\label{eq:functdensity}
\eea
Note that $\phi^a$ $(a=1,2)$ represents two components scalars.
\bea
\phi^a&=&\begin{pmatrix} \phi^a_1 \\ \phi^a_2 \end{pmatrix}.
\eea
We assume that the particle number violating term
$B^2$ turned on when the universe begins to expand at $x^0 = 0$. 
The initial value for the scale factor is $a_0$.
Since the Hamiltonian $H_0$ and the particle number $N$ 
commute with each other, the 
exponential factors in the grand canonical distribution function 
are factorized as shown in Eq.(\ref{eq:functdensity}).
$\hat{N}$ is a functional derivative acting 
on $\phi^1$ and corresponds to the number operator in Eq.(\ref{eq:N}).
\bea
\hat{N}=\int d^3 x a_0^3  j^0 = -i \int d^3 x  \left({\phi^1_2} \frac{\delta}{\delta \phi^1_1}-\phi^1_1 \frac{\delta}{\delta \phi^1_2}\right). 
\eea
We first investigate the functional representation for the
density matrix with zero chemical potential. 
\bea
\langle \phi^1 |\exp(-\beta H_0)| \phi^2 \rangle &=&  \int_{\phi(u=\beta)=\phi^1, \phi(u=0)=\phi^2} d \phi \exp(-S_E) \nn \\
&=& C_0 \exp(-S^{\mu=0}_{E cl}[\phi^1,\phi^2]).
\eea
where $S_E$ is a Euclidean action for the complex scalar field
and $S^{\mu=0}_{E cl}$ is the one for the classical trajectory with the
boundary conditions at the Euclidean time $u=0$ and $u=\beta$.
$C_0$ is a constant.
Explicitly $S_E$  is given as,
\bea
S_E=\int_{0}^{\beta} du \int d^3 x a_0^3
\Bigl{[}\frac{\partial \phi^\dagger}{\partial u}
\frac{\partial \phi}{\partial u}+
\frac{\nabla \phi^\dagger \cdot \nabla \phi}{a_0^2}+
m_\phi^2 \phi^\dagger \phi \Bigr{]}.
\eea
and $S^{\mu=0}_{E cl}$ becomes,
\bea
S^{\mu=0}_{E cl}[\phi^1,\phi^2]= -\frac{a_0^6}{2} \sum_{i,j=1,2} \int \frac{d^3 k}{(2\pi)^3}
\phi_i^b({\bf k})c^{ab} c^{cd} \kappa_{ij}^{0ac}(-{\bf k}) \phi_j^{d}(-{\bf k}).
\eea
$\kappa^0$ represents $\kappa$ defined in Eq.(\ref{eq:kappa}) for the zero chemical potential case. One can find,
\bea
\kappa_{ij}^{011}({\bf -k})&=&\kappa_{ij}^{022}({\bf -k})=-\frac{1}{a_0^3} 
\frac{\omega(k) \cosh \beta \omega(k)}{\sinh \beta \omega(k)} \delta_{ij}, 
\nn \\
\kappa_{ij}^{012}({\bf -k})&=&
\kappa_{ij}^{021}({\bf -k})=-\frac{1}{ a_0^3} \frac{\omega(k)}{\sinh \beta \omega(k)} \delta_{ij},
\eea
where $\omega(k)=\sqrt{\frac{k^2}{a_0^2}+m_\phi^2}$. To obtain the
functional representation of the density matrix for 
non-zero chemical potential, one notes the action of $\exp(\mu \beta \hat{N})$ generates 
$O(2)$ rotation among $\phi_1, \phi_2$ with a
complex angle $i \mu \beta$, 
\bea
\exp(\mu \beta \hat{N}) \begin{pmatrix} \phi^1_1 \\ \phi^1_2 \end{pmatrix}
=O(i \mu \beta) \begin{pmatrix} \phi^1_1 \\ \phi^1_2 \end{pmatrix},
\eea
where $O(i \mu \beta)$ is a rotation matrix,
\bea
O(i \mu \beta)=\begin{pmatrix} \cosh \mu \beta & -i \sinh \mu \beta \\
+ i \sinh \mu \beta & \cosh \mu \beta 
\end{pmatrix}.
\eea
Therefore the action of $\exp(\mu \beta \hat{N})$ replaces $\phi^1$
with $O(i \mu \beta) \phi^1$. The resulting functional representation of the
density matrix for non-zero chemical potential is,
\bea
<\phi^1|\exp(-\beta(H_0-\mu N))| \phi^2>&=& <O(i\mu \beta) \phi^1|\exp(-\beta H_0)|\phi^2> \nn \\
&\equiv &C \exp(-S_{cl}^{\mu}[\phi^1, \phi^2]),
\eea
where,
\bea
S^{\mu}_{cl}[\phi^1, \phi^2]&=&S^{\mu=0}_{cl}[O(i \beta \mu) \phi^1,\phi^2]
\nn \\
&=&
-\frac{a_0^6}{2} \sum_{i,j=1,2} \int \frac{d^3 k}{(2\pi)^3}
\phi_i^b({\bf k})c^{ab} c^{cd} \kappa_{ij}^{ac}(-{\bf k}) \phi_j^{d}(-{\bf k}).
\eea 
$\kappa$ for non-zero chemical potential is given as,
\bea
\kappa_{ij}^{11}(-{\bf k})&=&\kappa_{ij}^{22}(-{\bf k})=-\frac{1}{a_0^3} 
\frac{\omega(k) \cosh \beta \omega(k)}{\sinh \beta \omega(k)} \delta_{ij}, 
\nn \\
\kappa_{ij}^{12}({\bf -k})&=&-\frac{1}{a_0^3} \frac{\omega(k)}{\sinh \beta \omega(k)} O^{T}_{ij}(i \mu \beta), \nn \\
\kappa_{ij}^{21}({\bf -k})&=&-\frac{1}{a_0^3} \frac{\omega(k)}{\sinh \beta \omega(k)} O_{ij}(i \mu \beta).
\eea
The normalization factor $C$ can be determined by the condition ${\rm Tr}(\rho)=1$.
\bea
<\phi^1|\rho|\phi^2>=\frac{\exp(-S^{\mu}_{cl}[\phi^1,\phi^2])}
{\int 
d \phi_1 d \phi_2 \exp[-S^{\mu}_{cl}[\phi,\phi]]},
\label{eq:rho}
\eea
where,
\bea
S^{\mu}_{cl}[\phi,\phi]&=&a_0^3 \int \frac{d^3 k}{(2 \pi)^3}  \frac{\omega({\bf k})(\cosh \beta \omega({\bf k})-\cosh \beta \mu)}{\sinh \beta \omega(k)} 
\phi_i({\bf k})\phi_i(-{\bf k}) \nn \\
                      &=&\frac{1}{2} \int d^3 {\bf x} d^3 {\bf y} 
\phi_i({\bf x}) D({\bf x-y}) 
\phi_i({\bf y}), 
\eea
with $D({\bf r})$ defined as,
\bea
D({\bf r})=2a_0^3 \int \frac{d^3 k}{(2 \pi)^3}  \frac{\omega({\bf k})(\cosh \beta \omega({\bf k})-\cosh \beta \mu)}{\sinh \beta \omega(k)}  \exp(-i {\bf r \cdot k}).
\eea
The functional representation of 
the density matrix in Eq.(\ref{eq:rho}) is 
used for obtaining the initial condition of the Green functions which are 
needed to solve the differential equations of Eq.(\ref{eq:twodif}).
The Green function at $x^0=y^0=0$ is defined as,
\bea
G^{ab}_{ij}({\bf x}, x^0=0, {\bf y}, y^0=0)&=&
\rm{Tr}[\hat{\phi}_j({\bf y}) \hat{\phi}_i({\bf x}) \rho], \nn \\
&=& \frac{\int d\phi_1  d\phi_2 \phi_j ({\bf y}) 
\phi_i({\bf x}) \exp[-S^\mu_{cl}[\phi, \phi]]}{\int d\phi_1  d\phi_2 
 \exp[-S^\mu_{cl}[\phi, \phi]]},
\eea
and it can be computed with the generating functional,
\bea
W[J]&=& \frac{\int d\phi_1  d\phi_2  
\exp[-S^\mu_{cl}[\phi, \phi]+\int d^3 {\bf x}J_i({\bf x}) \phi_i({\bf x})]}{\int d\phi_1  d\phi_2 
 \exp[-S^\mu_{cl}[\phi, \phi]]}, \nn \\
    &=& \exp 
\Bigl{[}\frac{1}{2} \int d^3 x d^3 y J_i(x) D^{-1}(x-y) J_i(y) 
\Bigr{]}.
\eea
Differentiating $W[J]$ with the source term twice, one obtains,
\bea
G_{ij}^{ab}({\bf x},x^0=0,{\bf y},y^0=0)=\frac{\delta^2 W[J]}
{\delta J_i({\bf x}) \delta J_j({\bf y})}\Biggr{|}_{J=0}
=D^{-1}({\bf x-y}) \delta_{ij},
\label{eq:D}
\eea
where $D^{-1}({\bf x-y})$ satisfies
\bea
\int d^3 {\bf y} D({\bf x-y}) D^{-1}({\bf y-z})=\delta^3({\bf x-z}). 
\label{eq:Dinv}
\eea
The Fourier transformation of $D({\bf x-y})$ and its inverse 
$D^{-1}({\bf x-y})$
are,
\bea
D({\bf k})&=&2 a_0^3 \omega(k) 
\frac{\cosh \beta \omega(k)-\cosh \beta \mu}{\sinh \beta \omega(k)},\nn \\
D^{-1}({\bf k})
&=&\frac{1}{2 a_0^3 \omega(k)} 
\Bigl{[}
\frac{\sinh \beta \omega(k)}{\cosh \beta \omega(k)-\cosh \beta \mu}
\Bigr{]}.
\label{eq:FD}
\eea
Next we define the Fourier transform of the Green functions,
\bea
G^{ab}_{ij}(x,y)= \int \frac{d^3 k}{(2 \pi)^3}
G^{ab}_{ij}(x^0,y^0,{\bf k}) e^{ -i {\bf k \cdot x}}.
\eea
Using Eq.(\ref{eq:Dinv}) and Eq.(\ref{eq:FD}),
we obtain the initial value of
the Fourier transformation of the Green function, 
\bea
G^{ab}_{ij}(x^0=0, y^0=0,{\bf k})&=& \delta_{ij}\frac{1}{D({\bf k})},\nn \\
                                 &=&\delta_{ij}
\frac{1}{2 \omega({\bf k}) a_0^3}
\Bigl{[}\frac{\sinh \beta \omega({\bf k})}{\cosh \beta \omega({\bf k})-\cosh \beta \mu} \Bigr{]}.
\label{eq:initialcon}
\eea
Since we obtain the initial condition of Green
function, 
one can use it to solve the Schwinger Dyson equations.

In Friedman Robertson Walker metric, the Laplacian is given as,
\bea
\nabla_\mu \nabla^\mu &=&
\frac{\partial^2}{{\partial x^0}^2}-\frac{1}{a(x^0)^2}
\nabla \cdot \nabla+ 3 \frac{\dot{a}}{a} \frac{\partial}{\partial x^0}
\eea
Therefore, the Fourier transformation of Green functions satisfy,
\bea
&&(\frac{\partial^2 }{\partial x^{02}}+\frac{{\bf k}^2}{a(x^0)^2}+\tilde{m}_m(x^0)^2
+ 3 H \frac{\partial }{\partial x^{0}}) G^{ab}_{mn}(x^0,y^0, {\bf k}) \nn \\
&=&-i\frac{c^{ab}}{a(x^0)^3} \delta(x^0-y^0) \delta_{mn}
-i \delta(x^0) a_0^3 \kappa^{ac}_{ml}({\bf k}) c^{cd} G_{ln}^{db}(0,y^0,{\bf k}) , \nn \\
&&(\frac{\partial^2 }{\partial y^{02}}+\frac{{\bf k}^2}{a(y^0)^2}+
\tilde{m}_n(y^0)^2
+ 3 H \frac{\partial }{\partial y^{0}}) G^{ab}_{mn}(x^0,y^0, {\bf k})\nn \\
&=&-i\frac{c^{ab}}{a(y^0)^3} \delta(x^0-y^0) \delta_{mn}
-i\delta(y^0)a_0^3 G_{ml}^{ac}(x^0,0,{\bf k})c^{cd}
\kappa_{ln}^{db}({\bf k}),
\label{eq:diffgreen}
\eea
%

To solve Eq.(\ref{eq:diffgreen}), we introduce $\hat{G}$ through the 
following equantion.
\bea
G^{ab}_{mn}(x^0,y^0,k)=\left(\frac{a_0}{a(x^0)}\frac{a_0}{a(y^0)}
\right)^{\frac{3}{2}}\hat{G}^{ab}_{mn}(x^0,y^0, k).
\eea
The differential equations are rewritten as,
\bea
&&(\frac{\partial^2 }{\partial x^{02}}+\frac{{\bf k}^2}{a(x^0)^2}+
\overline{m}_m(x^0)^2) \hat{G}^{ab}_{mn}(x^0,y^0, {\bf k}) \nn \\
&=&-i\frac{c^{ab}}{a_0^3} \delta(x^0-y^0) \delta_{mn}
-i \delta(x^0) a_0^3 \kappa^{ac}_{ml}({\bf k}) c^{cd} \hat{G}_{ln}^{db}(0,y^0,{\bf k}), 
\label{eq:x0} \\
&&(\frac{\partial^2 }{\partial y^{02}}+\frac{{\bf k}^2}{a(y^0)^2}+
\overline{m}_n(y^0)^2
) \hat{G}^{ab}_{mn}(x^0,y^0, {\bf k})\nn \\
&=&-i \frac{c^{ab}}{a_0^3} \delta(x^0-y^0) \delta_{mn}
-i\delta(y^0)a_0^3 \hat{G}_{ml}^{ac}(x^0,0,{\bf k})c^{cd}
\kappa_{ln}^{db}({\bf k}),
\label{eq:y0}
\eea
where $\overline{m}_m(x^0)^2=\tilde{m}_m(x^0)^2-\frac{9 H^2}{4}-
\frac{3}{2} \frac{dH}{dx^0}.$
In the following, we denote two independent solutions of the homogeneous
differential equation of Eq.(\ref{eq:x0}) as $f_m(x^0)$ and $g_m(x^0)$.
\bea
(\frac{\partial^2 }{\partial x^{02}}+\frac{{\bf k}^2}{a(x^0)^2}+
\overline{m}_m(x^0)^2) \Bigl{\{} \begin{array}{c} 
f_m(x^0) =0, \\ 
g_m(x^0) =0.
\end{array}
\label{eq:fg}
\eea
To solve the differential equations for Green functions,  we introduce 
the following four by four matrices.
\bea
\hat{G}(x^0, y^0, {\bf k})=\begin{pmatrix}
\hat{G}_{11}(x^0,y^0,{\bf k}) & \hat{G}_{12}(x^0,y^0,{\bf k}) \\
\hat{G}_{21}(x^0,y^0,{\bf k}) & \hat{G}_{22}(x^0,y^0,{\bf k})
\end{pmatrix}.
\eea
where each $\hat{G}_{ij}(x^0,y^0, {\bf k})$ is given by 
a two by two matrix.
\bea
\hat{G}_{ij}(x^0,y^0, {\bf k})=\begin{pmatrix} 
\hat{G}_{ij}^{11}(x^0,y^0, {\bf k}) & \hat{G}_{ij}^{12}(x^0,y^0, {\bf k}) \\
\hat{G}_{ij}^{21}(x^0,y^0, {\bf k}) & \hat{G}_{ij}^{22}(x^0,y^0, {\bf k})
\end{pmatrix}.
\eea
In this notation, $c$ and $\kappa$ are given as,
\bea
c=\begin{pmatrix} 1 & 0 & 0 &0 \\
                  0 & -1 & 0& 0 \\
                  0 & 0 & 1 & 0 \\
                  0 & 0 & 0 & -1 \end{pmatrix}, \quad 
\kappa= \begin{pmatrix} \kappa_{11}(-{\bf k}) & \kappa_{12}(-{\bf k}) \\
                                  \kappa_{21}(-{\bf k}) & 
\kappa_{22}(-{\bf k}) \end{pmatrix}.
\eea
where each $\kappa_{ij}(-{\bf k})$ is a two by two matrix and is given by,
\bea
\kappa_{ij}(-{\bf k})=\begin{pmatrix} \kappa_{ij}^{11}(-{\bf k})
& \kappa_{ij}^{12}(-{\bf k})  \\
 \kappa_{ij}^{21}(-{\bf k}) & \kappa_{ij}^{22}(-{\bf k})
\end{pmatrix}.
\eea
\def\hG{\hat{G}}
Now let us solve Eq.(\ref{eq:x0}) and Eq.(\ref{eq:y0}). 
When $x^0 > y^0$, one first writes $\hG(x^0,y^0)$ in terms of $\hG(x^0,0)$
and $\frac{\partial \hG(x^0,y^0)}{\partial y^0}\Bigl{|}_{y^0=0}$.
\bea
\hG_{mn}^{ab}(x^0,y^0, {\bf k})&=&
\hG_{mn}^{ab}(x^0,0, {\bf k}) \omega_n(y^0)+ \frac{\partial
\hG_{mn}^{ab}(x^0,y^0, {\bf k})}{\partial y^0}\Bigr{|}_{y^0=0} z_n(y^0),
\label{eq:G1}  \\
\frac{\partial \hG_{mn}^{ab}(x^0,y^0, {\bf k})}{\partial y^0}\Bigr{|}_{y^0=0}
&=&-i a_0^3 \hG_{ml}^{ac}(x^0,0,{\bf k}) c^{cd}\kappa^{db}_{ln}(-{\bf k}),
\label{eq:G2}
\eea
Next we write $G(x^0,0)$ with $G(0,0)$ as,
\bea
\hG_{mn}^{ab}(x^0,0,{\bf k})&=& \omega_m(x^0) 
\hG_{mn}^{ab}(0,0,{\bf k}) + z_m(x^0) 
\frac{\partial \hG_{mn}^{ab}(x^0,0,{\bf k})}{\partial x^0}\Bigr{|}_{x^0=0}
,
\label{eq:G3}
 \\
\frac{\partial \hG_{mn}^{ab}(x^0,0,{\bf k})}{\partial x^0}\Bigr{|}_{x^0=0}
&=& -i\frac{c^{ab}}{a_0^3} \delta_{mn}-i a_0^3 \kappa^{ac}_{ml}(-{\bf k})
c^{cd} G^{db}_{ln}(0,0,{\bf k}),
\label{eq:G4}
\eea
where $w_n(x^0)$ and $z_n(x^0)$ are defined as,
\bea
w_n(x^0)&=&
\frac{f_n(x^0) \dot{g}_n(0)-g_n(x^0) \dot{f}_n(0)}
{f_n(0) \dot{g}_n(0)-g_n(0) \dot{f}_n(0)},
 \nn \\
z_n(x^0)&=&
\frac{-f_n(x^0) g_n(0)+g_n(x^0) f_n(0)}
{f_n(0) \dot{g}_n(0)-g_n(0) \dot{f}_n(0)}.
\label{eq:wz}
\eea
Using Eqs.(\ref{eq:G1}-\ref{eq:G4}), one can write $\hG(x^0, y^0)$ in terms
of $\hG(0,0)$ where $\hG(0,0)$ is obtained in Eq.(\ref{eq:initialcon}) 
in the previous section.
To compute all components of $\hG$, one introduces the diagonal matrices $w(x^0)$ and $z(x^0)$,
\bea
w(x^0)&=& \begin{pmatrix} w_1(x^0) & 0 & 0 & 0 \\
                        0 & w_1(x^0) & 0 & 0 \\
                        0 & 0 & w_2(x^0) & 0 \\
                        0 & 0 & 0 & w_2(x^0)  \end{pmatrix},  \\
z(x^0)&=&\begin{pmatrix} z_1(x^0) & 0 & 0 & 0 \\
                        0 & z_1(x^0) & 0 & 0 \\
                        0 & 0 & z_2(x^0) & 0 \\
                        0 & 0 & 0 & z_2(x^0)  \end{pmatrix}.
\eea
Using them, one can write the solution $\hat{G}(x^0, y^0)$ for $x^0> y^0$ as, 
\bea
\hG(x^0,y^0,{\bf k})&=&(w(x^0)-z(x^0) ia_0^3 \kappa c) G(0,0,{\bf k})
(w(y^0)-i c \kappa a_0^3 z(y^0)) \nn \\
&-&i z(x^0) 
\frac{c}{a_0^3} (w(y^0)-i c \kappa a_0^3 z(y^0)).
\label{eq:matrixform}
\eea
For $x^0 < y^0$, one can also write 
the solution in the matrix form  similar to 
Eq.(\ref{eq:matrixform}). The result is,
\bea
\hG(x^0,y^0,{\bf k})&=&(w(x^0)-z(x^0) ia_0^3 \kappa c) G(0,0,{\bf k})
(w(y^0)-i c \kappa a_0^3 z(y^0))\nn \\
&-&i(w(x^0)-z(x^0) i a_0^3 \kappa c) \frac{c}{a_0^3} z(y^0).
\label{eq:matrixform2}
\eea
By combining Eq.(\ref{eq:matrixform}) with Eq.(\ref{eq:matrixform2}),
one obtains,  
\bea
G(x^0,y^0,{\bf k})&=&
\left(\frac{a_0}{a(x^0)}\frac{a_0}{a(y^0)}
\right)^{\frac{3}{2}}
\Bigl{[} (w(x^0)-z(x^0) ia_0^3 \kappa c) G(0,0,{\bf k})
(w(y^0)-i c \kappa a_0^3 z(y^0)) -z(x^0) \kappa z(y^0) \nn \\
&-&i \theta(x^0-y^0)z(x^0) \frac{c}{a_0^3} \omega(y^0)  
-i \theta(y^0-x^0) \omega(x^0) \frac{c}{a_0^3} z(y^0) \Bigr{]}.
\eea

Now we are ready to write all the Green functions explicitly.
The diagonal elements, $\hat{G}_{ii}$ $(i=1,2)$ are  given as,
\bea
\hat{G}_{ii}(x^0,y^0,k)&=&
\frac{(w_i(x^0) w_i(y^0)+ \omega(k)^2 z_i(x^0) z_i(y^0)) \sinh \beta \omega(k)}
{ 2 a_0^3 \omega(k)(\cosh \beta \omega(k)-\cosh \beta \mu)} \begin{pmatrix} 1 & 1 \\
                                                                     1 & 1 \end{pmatrix} \nn \\
&+& \frac{i}{2 a_0^3} (w_i(x^0) z_i(y^0)-z_i(x^0) w_i(y^0)) 
\begin{pmatrix} \epsilon(x^0-y^0) & -1 \\
                     1   & -\epsilon(y^0-x^0) \end{pmatrix},
\eea
where 
$\epsilon(x^0-y^0)=\theta(x^0-y^0)-\theta(y^0-x^0)$. 
The off-diagonal ones, $\hat{G}_{ij} (i \ne j)$ are given by,
\bea
\hat{G}_{12}(x^0,y^0,k)&=&\frac{\sinh \beta \mu}{2 a_0^3 (\cosh \beta \omega(k)-\cosh \beta \mu)}
(-z_2(y^0) w_1(x^0)+w_2(y^0) z_1(x^0)) \begin{pmatrix} 1 & 1 \\ 1 & 1 \end{pmatrix}, \nn \\ 
\hat{G}_{21}(x^0,y^0,k)&=&\frac{\sinh \beta \mu}{2 a_0^3(\cosh \beta \omega(k)-\cosh \beta \mu)}
(-z_2(x^0) w_1(y^0)+w_2(x^0) z_1(y^0))
\begin{pmatrix} 1 & 1 \\ 1 & 1 \end{pmatrix}. \nn \\
\eea
We also write $G_{12}^{12}$ explicitly,
\bea
G_{12}^{12}(x^0,y^0,{\bf k})=\left(\frac{a_0}{a(x^0)}\frac{a_0}{a(y^0)}
\right)^{\frac{3}{2}}
\frac{\sinh \beta \mu}{2 a_0^3
(\cosh \beta \omega(k)-\cosh \beta \mu)}
(-z_2(y^0) w_1(x^0)+w_2(y^0) z_1(x^0)).\nn \\
\eea
Using the result, one can write the current density,
\bea
<j_0(x^0)>&=& \left(\frac{a_0}{a(x^0)}\right)^3 \int \frac{d^3 k}{(2 \pi)^3}
\frac{\sinh \beta \mu}{2 a_0^3 (\cosh \beta \omega(k)-\cosh \beta \mu)}\nn \\
&&[-\dot{w}_1(x^0)z_2(x^0)+w_1(x^0) \dot{z}_2(x^0)-\dot{w}_2(x^0) z_1(x^0)
+w_2(x^0) \dot{z}_1(x^0)
].
\label{eq:density}
\eea
Now let us examine the solutions of homogeneous differential equations
of Eq.(\ref{eq:fg}) for the case that the scale factor grows exponentially with respect to time, 
\bea
a(x^0)=a_0 \exp(H x^0).
\label{eq:exp}
\eea
The masses $\tilde{m}_m$ in Eq.(\ref{eq:mtilde}) are independent of time
and an analytic form for the Green functions can be obtained. 
In this case one can introduce the conformal time \cite{CalzettaHu}, 
\bea
\eta=-\frac{k}{H a(x^0)}=-\frac{k}{H e^{H x^0}}. 
\eea
where we can set $a(x_0=0)=a_0=1$ without loss of generality. 
One finds the $f_m$ and $g_m$ ($m=1,2$) satisfy the differential equation 
for Bessel function,
\bea
\Bigr{[}\frac{\partial^2 }{\partial \eta^2}+\frac{1}{\eta} \frac{\partial
}{\partial \eta}+1+\frac{\rho_m^2}{\eta^2}
\Bigl{]} \Bigr{\{} 
\begin{array}{c}
f_m(\eta)=0 \\ g_m(\eta)=0,
\end{array}
\label{eq:homo}
\eea 
where $\rho_m$ ($m=1,2$) is given as, 
\bea
\rho_m=\frac{\overline{m}_m}{H}, 
\eea
\bea
\rho_1&=&
\sqrt{\frac{m_\phi^2}{H^2}-\frac{B^2}{H^2}-
12(\alpha_3+\alpha_2)-\frac{9}{4}} 
,\nn \\
\rho_2&=& \sqrt{\frac{m_\phi^2}{H^2}+\frac{B^2}{H^2}-
12(\alpha_3-\alpha_2)-\frac{9}{4}} 
.
\label{eq:rho2}
\eea
One can choose the Hankel function $H_{i \rho_m}$
as one of the solution.
\bea
f_{m}(x^0)&=&H_{i \rho_m} [\eta]= \frac{1}{\sinh \rho_m \pi}
\Big{(} e^{\rho_m \pi } J_{i \rho_m}[\eta] - J_{-i \rho_m}[\eta] 
\Big{)},
\label{eq:Han}
\eea 
where we also show the formula which relates the Hankel function to
the Bessel function.
The current density of Eq.(\ref{eq:density}) is also written
in terms of the derivative with conformal time.
Using the relation of the derivatives,
\bea
\frac{\partial}{\partial x^0}=\frac{k}{a(x^0)} 
\frac{\partial}{\partial \eta}, 
\eea
one can write $w_n$ and $z_n$ in Eq.(\ref{eq:wz}),
\bea
w_n(x^0)&=&\frac{H_{i \rho_n}[\eta] H_{i \rho_n}^{\prime \ast}[\eta_0]-
H_{i \rho_n}^{\ast}[\eta] H_{i \rho_n}^{\prime}[\eta_0]}
{H_{i \rho_n}[\eta_0] H_{i \rho_n}^{\prime \ast}[\eta_0]-
H_{i \rho_n}^{\ast}[\eta_0] H_{i \rho_n}^{\prime}[\eta_0]},\nn \\
z_n(x^0)&=& \frac{-H_{i \rho_n}[\eta] H_{i \rho_n}^{\ast}[\eta_0]+
H_{i \rho_n}^{\ast}[\eta] H_{i \rho_n}[\eta_0]}
{k(H_{i \rho_n}[\eta_0] H_{i \rho_n}^{\prime \ast}[\eta_0]-
H_{i \rho_n}^{\prime}[\eta_0] H^{\ast}_{i \rho_n}[\eta_0])},
\eea
where $H^{\prime} \equiv \frac{\partial H}{ \partial \eta}$
and $\eta_0=-\frac{k}{H}$.
The time derivatives of $w_n$ and $z_n$ are also written with the
derivatives with the conformal time.
\bea
\dot{w}_n(x^0)&=&\frac{k}{a(x^0)}
\frac{H_{i \rho_n}^\prime[\eta] H_{i \rho_n}^{\prime \ast}[\eta_0]-
H_{i \rho_n}^{\prime \ast}[\eta] H_{i \rho_n}^{\prime}[\eta_0]}
{H_{i \rho_n}[\eta_0] H_{i \rho_n}^{\prime \ast}[\eta_0]-
H_{i \rho_n}^{\ast}[\eta_0] H_{i \rho_n}^{\prime}[\eta_0]}, \nn \\
\dot{z}_n(x^0)&=& \frac{1}{a(x^0)}
\frac{-H^\prime_{i \rho_n}[\eta] H_{i \rho_n}^{\ast}[\eta_0]+
H_{i \rho_n}^{\ast \prime}[\eta] H_{i \rho_n}[\eta_0]}
{(H_{i \rho_n}[\eta_0] H_{i \rho_n}^{\prime \ast}[\eta_0]-
H_{i \rho_n}^{\prime}[\eta_0] H^{\ast}_{i \rho_n}[\eta_0])}. 
\eea
Furthermore, we introduce $W_n$ and $Z_n$ as functions of the conformal 
time.
\bea
W_n[\eta, \eta_0]&=&H_{i \rho_n}[\eta] H_{i \rho_n}^{\prime \ast}[\eta_0]-
H_{i \rho_n}^{\ast}[\eta] H_{i \rho_n}^{\prime}[\eta_0], \nn \\
Z_n[\eta, \eta_0]&=&-H_{i \rho_n}[\eta] H_{i \rho_n}^{\ast}[\eta_0]+
H_{i \rho_n}^{\ast}[\eta] H_{i \rho_n}[\eta_0].
\label{eq:WZ}
\eea
With these formulas, one can write the current density of 
Eq.(\ref{eq:density}) with the
conformal time.
\bea
&&<j_0(x^0)>= e^{-4 H x^0}\int \frac{d^3 k}{(2 \pi)^3}
\frac{\sinh \beta \mu}{ (\cosh \beta \omega(k)-\cosh \beta \mu)}\nn \\
&&\Bigl{[}\frac{-{W^\prime}_1[\eta, \eta_0] Z_2[\eta, \eta_0]+W_1[\eta, \eta_0]
 {Z_2}^\prime[\eta, \eta_0] -{W_2}^\prime[\eta, \eta_0] Z_1[\eta, \eta_0]
+W_2[\eta,\eta_0] {Z_1}^\prime[\eta, \eta_0]}{2 W_1[\eta_0,\eta_0] W_2[\eta_0,\eta_0]}
\Bigr{]}. \nn \\
\label{eq:current2}
\eea

Below we investigate some extreme limit of the particle number density of
Eq.(\ref{eq:density}) and Eq.(\ref{eq:current2}).
We first study the small limit of the Hubble parameter $H$ of 
Eq.(\ref{eq:current2}). 
When $H$ is small, $\rho_i=\sqrt{\frac{m_i^2}{H^2}}$ and $x=\frac{k}{H}$ become large. In appendix A, we derive the approximate formula for the small H limit.
From Eq.(\ref{eq:asmallimit}), 
one obtains the current density for small $H$ limit, 
\bea
&&<j_0(x^0)>= e^{-3 H x^0}\int \frac{d^3 k}{(2 \pi)^3}
\frac{\sinh \beta \mu}{ (\cosh \beta \omega(k)-\cosh \beta \mu)}\nn \\
&& \frac{1}{2}\Bigl{[}\Bigl{\{}\left(
\frac{\omega_1 \omega_1(x^0)}{\omega_2 \omega_2(x^0)}\right)^{\frac{1}{2}}+
\left(\frac{\omega_2 \omega_2(x^0)}{\omega_1 \omega_1(x^0)}
\right)^{\frac{1}{2}}
\Bigr{\}} 
\sin x(f(a, \sigma_1)-f(1, \sigma_1)) \sin x (f(a, \sigma_2)-f(1, \sigma_2)) \nn\\
&+&\Bigl{\{}
\left(\frac{\omega_1 \omega_2(x^0)}{\omega_2 \omega_1(x^0)}\right)^{\frac{1}{2}}+
\left(\frac{\omega_2 \omega_1(x^0)}{\omega_1 \omega_2(x^0)}
\right)^{\frac{1}{2}}
\Bigr{\}} 
\cos x(f(a, \sigma_1)-f(1, \sigma_1)) \cos x (f(a, \sigma_2)-f(1, \sigma_2)) 
\Bigr{]}, \nn \\
\eea 
where $\sigma_i=\frac{m_i}{k}$, $\omega_i=\sqrt{k^2+m_i^2}$ and $\omega_i(x^0)=\sqrt{\frac{k^2}{a(x^0)^2}+m_i^2}$. $f(a, \sigma)=\sqrt{\frac{1}{a^2}+\sigma^2}
-\sigma \sinh^{-1}(\sigma a) $.
One can also take the limit that the expansion rate $H$ vanishes.
(See Eq.(\ref{eq:azerolimit}).)
\bea
\lim_{H \rightarrow 0}<j_0(x^0)>
&=& \int \frac{d^3 k}{(2 \pi)^3}
\frac{\sinh \beta \mu}{ (\cosh \beta \omega(k)-\cosh \beta \mu)}\nn \\
&& \Bigl{[} \cos (\omega_1-\omega_2) x^0+\frac{1}{2}\Bigl{\{}
\frac{\omega_1}{\omega_2}+
\frac{\omega_2}{\omega_1}
-2\Bigr{\}} 
\sin \omega_1 x^0  \sin \omega_2 x^0
 \Bigr{]}. \nn \\
\label{eq:zerolimit}
\eea
Another interesting limit is the case 
that particle number violating mass term $B$
and $\alpha_2$ vanish. In this case, the particle number density per unit comoving volume is conserved. Since the comoving volume grows as 
$a(x^0)^3$, the density
in a unit physical volume 
decreases as $a(x^0)^{-3}$.
In appendix B, we explicitly derive the current density for the case with
{\bf $B=\alpha_2=0$} and the result is given as follows, 
\bea
\lim_{B \rightarrow 0, \alpha_2 \rightarrow 0}<j_0(x^0)>
&=& \left(\frac{a_0}{a(x^0)}\right)^3 \int \frac{d^3 k}{(2 \pi)^3}
\frac{\sinh \beta \mu}{ (\cosh \beta \omega(k)-\cosh \beta \mu)}.
\label{eq:bzerolimit}
\eea
\section{numerical results}
So far we take various limits and derive the corresponding formulae.
In this section, we study the exact formulae and the time dependence
of the current density numerically.  
We focus on the case that the scale factor grows exponentially 
with respect to time. Since the mode functions 
satisfy  Eq.(\ref{eq:homo}), the effective masses; 
$\bar{m}_1$ and $\bar{m}_2$ are given as follows, 
\begin{eqnarray}
  \bar{m}_{1}^{2} & = & m_{\phi}^{2} -B^{2} -12 (\alpha_{3}+\frac{3}{16}) H^{2} -12 \alpha_{2} H^{2} \label{eq:mb1}\\
 \bar{m}_2^2 & = & m_{\phi}^{2} + B^{2} -12 (\alpha_{3}+\frac{3}{16}) H^{2} 
- 12 \alpha_{2} H^{2} \label{eq:mb2}.
\end{eqnarray}
When the expansion rate $H$ is small and
the condition on $\alpha_i$ ($i=2,3$), 
\bea
\frac{|B|}{H} \gg \sqrt{|12 \alpha_2|},\quad \sqrt{12|\alpha_3+\frac{3}{16}|} 
\eea 
is satisfied,   
the leading contribution to the effective masses is given by the one 
of the flat space-time case specified with $B$ and $m_{\phi}$. 
We study the case that the effective masses are the same as those in flat 
space-time case and the H dependent parts of Eq.(\ref{eq:mb1}) and Eq.(\ref{eq:mb2})
vanish.
\begin{eqnarray}
 \alpha_{3} = \frac{-3}{16} , \hspace{1em} \alpha_{2} =0.
\label{eq:alphas}
\end{eqnarray}
Limiting ourselves to this case,
$\rho_{m}$ $(m=1,2)$ in Eq.(\ref{eq:rho2}) can be determined by the
coefficient of the particle number violating term $B$,
\begin{eqnarray}
  \rho_{1} = \frac{\sqrt{m_{\phi}^{2} -B^{2}}}{H} , \hspace{1em} \rho_{2} =
  \frac{\sqrt{m_{\phi}^{2} +B^{2}}}{H} . 
\end{eqnarray}
Without loss of generality, the initial value of the scale factor $a_0$
can be set to unity.
One can write the current density Eq.(\ref{eq:current2}) as,
\bea
<j_0(X^0)>=\int \frac{d^3 k}{(2\pi)^3} h(\mathbf{k},\mu,T) 
V (\eta , \eta_0,k),
\eea
where $h$ denotes the momentum distribution for the current density
and $V$ denotes the time evolution factor.  They are defined respectively as 
follows,
\bea
h(\mathbf{k},\mu,T) &=&\frac{\sinh \beta \mu}{\cosh \beta \omega - \cosh \beta \mu}, \\
V(\eta , \eta_0, k)&=&e^{-4HX^0}\frac{ -{W^\prime}_1[\eta, \eta_0] Z_2[\eta, \eta_0]+W_1[\eta, \eta_0] {Z_2}^\prime[\eta, \eta_0] + \Big{(} 1 \leftrightarrow 2 \Big{)} }{2W_1[\eta_0 ,\eta_0] W_2 [\eta_0 , \eta_0 ]}.
\eea
Because the time evolution factor $V$ is 
unity at the initial time,
one notes that the initial current density is simply given as,
\bea
<j_0(X^0=0)>=\int 
\frac{d^3 k}{(2\pi)^3} h(\mathbf{k},\mu,T).
\eea
\begin{figure}[htbp]
\begin{center}
\includegraphics[height=6.5cm,width=7cm,clip]{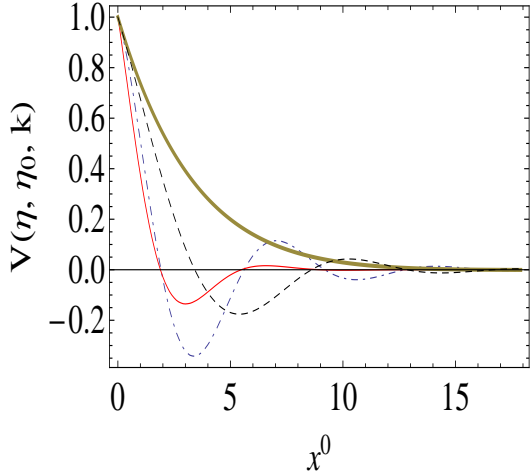}
\caption{The plot shows the time dependent factor V($\eta,\eta_0,k$) for 
different momentum $k$,  the expansion rate $H$ and particle number violating 
mass $B$. 
The chemical potential and
the temperature are fixed as $(\mu,T)=(5,20)$.
The dash-dotted line shows the result for the case $(B, H, k)=(3, 0.1, 5)$, the solid line shows the result for the case $(3, 0.2, 5)$ and the thick solid line shows the result for the case $(1, 0.1, 5 )$. 
The dashed line shows the result for the case  with  
$(B, H, k)=(3, 0.1, 20)$.}
\label{fig1}
\end{center}
\end{figure}
In Fig.1, we show the time evolution factor $V(\eta, \eta_0, k)$.
The period of oscillation tends to be long for the case that the mass squared
difference 
$2B^2$ is small and the momentum k is large.  
The damping speed becomes faster as the expansion rate $H$ is larger.
In Figs. 2 and 3, we show the momentum distribution function 
$h(k, \mu , T)$ for cases with different values of chemical potential $\mu$.
In  Fig. 2,  the case for $m_\phi > \mu$ is shown with $m_\phi=10$.
Fig. 3  shows the opposite case, i.e., $m_\phi < \mu$. Behavior of the two cases is different to each other because for the latter case, $h(k, \mu, T)$ has a pole at the momentum satisfying $\mu = \omega(k)$. 
We also find that for very large momentum compared with the temperature $T$
and the chemical potential $\mu$,
$h(k, \mu, T)$ becomes very small. Therefore one can set the upper limit of the momentum integration with a certain large momentum $k_{\max}$ and one can carry out the momentum integration approximately. 
\begin{figure}[htbp]
\begin{center}
\begin{tabular}{lccr}
\begin{minipage}{0.5\hsize}
\includegraphics[height=5.5cm,width=6.0cm,clip]{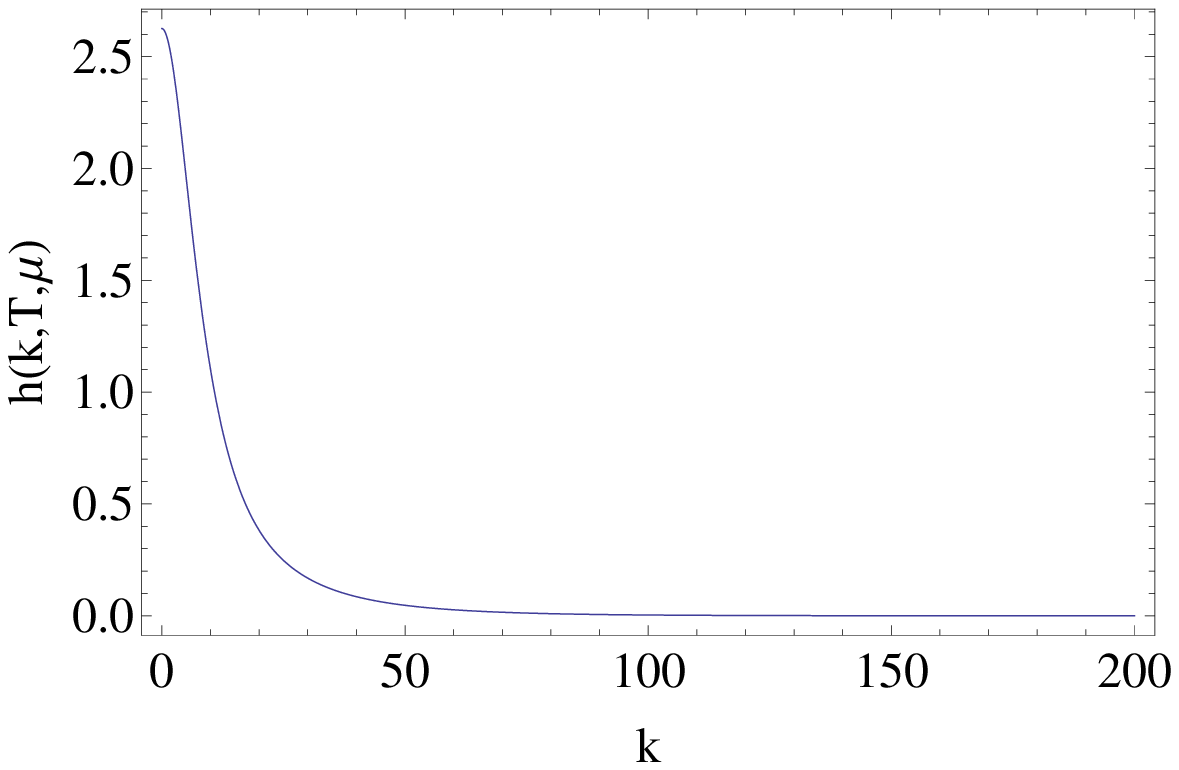}
\caption{\footnotesize{Momentum dependence of the distribution function h($k,\mu,T$). We choose parameters $(\mu,T)=(5,20)$.}}
\label{fig2}
\end{minipage}
&\qquad&\qquad&
\begin{minipage}{0.5\hsize}
\includegraphics[height=5.5cm,width=6.0cm,clip]{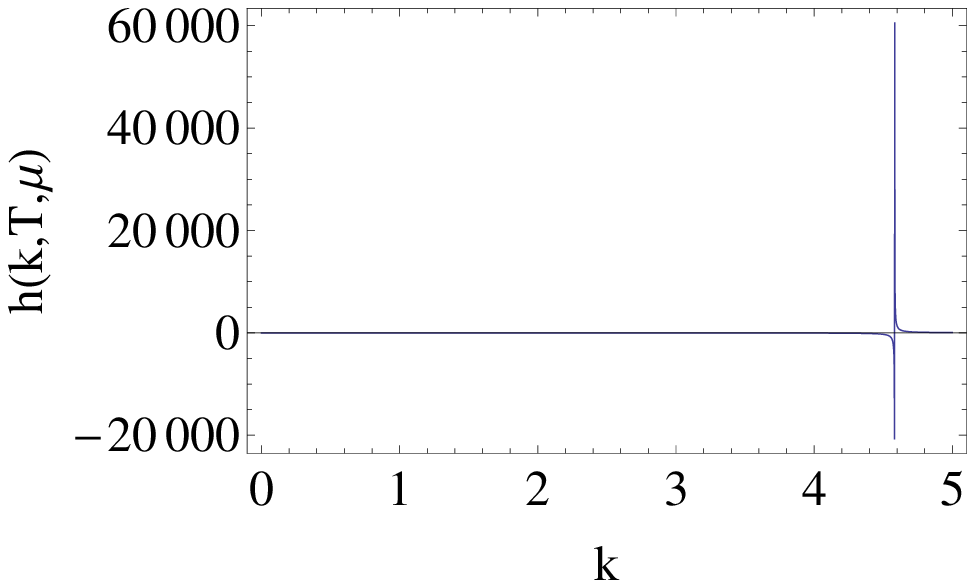}
\caption{\footnotesize{Momentum dependence of the distribution function h($k,\mu,T$).
 We choose parameters $(\mu, T)=(11,20)$.}}
\label{fig3}
\end{minipage}
\end{tabular}
\end{center}
\end{figure}
%

Below, we carry out momentum integration and show
the time variation of the current density.
We set the upper limit of the momentum integration $k_{max}=200$.
 We show the parameter dependence for time evolution of the current density in 
Fig.4 $\sim$ Fig.8.
First, we show the dependence on the expansion rate $H$ in Fig. 4 and Fig. 5. 
The expansion rate affects the damping speed of the current density. 
In fact, as the expansion rate $H$ becomes larger, 
the damping speed is faster.  We notice that the current density is suppressed 
even for the case that the expansion rate $H$ vanishes.
This is clearly seen from the behavior of the thick solid line of Fig. 5.
As shown in Fig. 1, the period of oscillation in $V$ 
varies depending on momentum $k$.
Therefore, contributions from different k interfere destructively and their sum
becomes small.
In Fig.6, we plot $B$ dependence for time evolution of the current density. 
The period of oscillation becomes shorter as the mass squared difference
is larger.
\begin{figure}[bhtp]
\begin{center}
\begin{tabular}{lcccr}
\begin{minipage}{0.32\hsize}
\includegraphics[height=4.7cm,width=4.7cm]{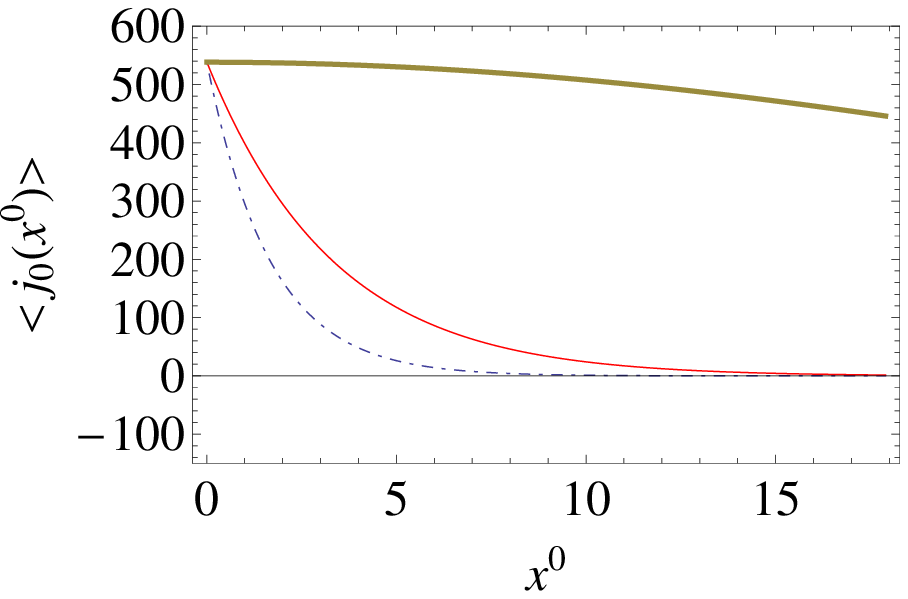}
\caption{\scriptsize{Curvature (H)  effect on time evolution of current density. 
The dash-dotted line, the solid line and the thick solid line show the case
for  $H=0.2,0.1,$ and $0$
respectively. ($B,\mu,T$)=($1,5,20$)
for all the lines.}}
\label{fig4}
\end{minipage}
&\quad &
\begin{minipage}{0.32\hsize}
\includegraphics[height=4.7cm,width=4.7cm]{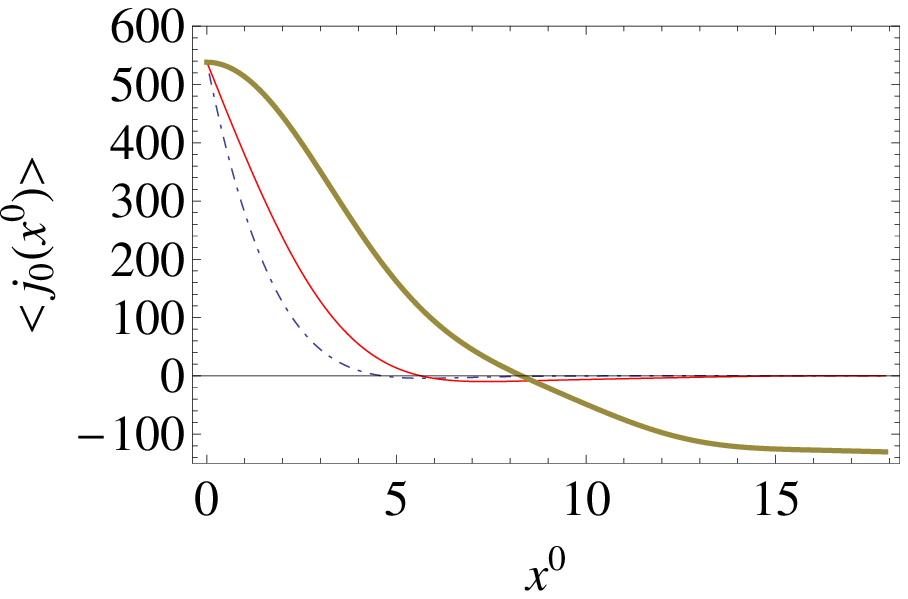}
\caption{\scriptsize{Curvature (H)  effect on time evolution of current density.
The dash-dotted line, the solid line and the thick solid line show the case for  $H=0.2,0.1,$ and $0$
respectively.
($B,\mu,T$)=($3,5,20$)
for all the lines.}}
\end{minipage}
&\quad & \begin{minipage}{0.32\hsize}
\includegraphics[height=4.7cm,width=4.7cm,clip]{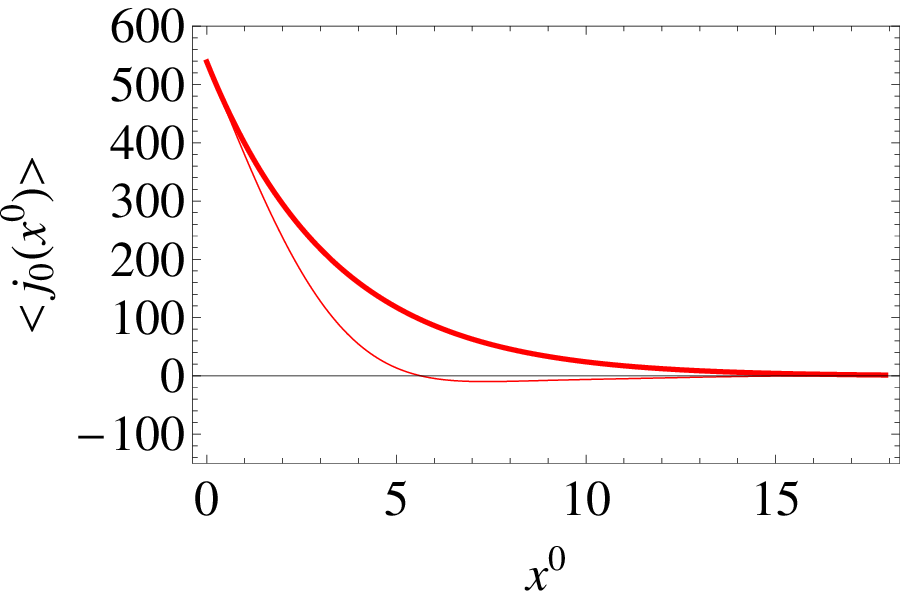}
\caption{
\scriptsize{The mass squared difference ($m_2^2-m_1^2=2 B^2$) dependence    
of current density. The solid line and the thick solid line
show the case $B=3$ and $B=1$ respectively.
($H,\mu,T$)=($0.1,5,20$) for all the lines.}} 
\end{minipage}
\end{tabular}
\end{center}
\end{figure}
\begin{figure}[htbp]
\begin{center}
\begin{tabular}{lccr}
\begin{minipage}{0.5\hsize}
\includegraphics[height=6.5cm,width=6.5cm,clip]{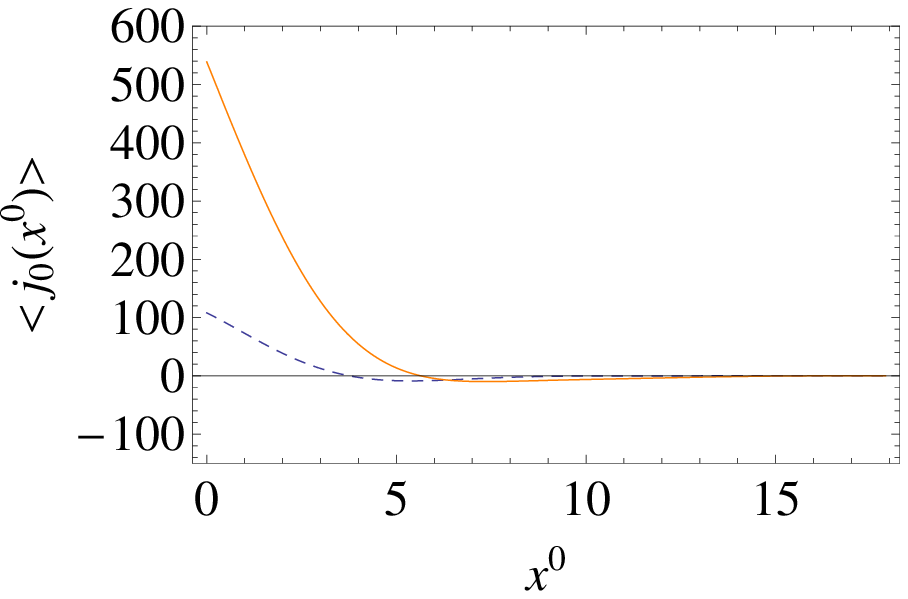}
\caption{\footnotesize{Dependence on temperature $T$ of time evolution of current density.
The dashed line and the solid line show the case $T=10$ and $T=20$
respectively.  $(B,H,\mu)=(3,0.1,5)$
for all the lines.}}
\label{fig7}
\end{minipage}
&\qquad& \qquad &
\begin{minipage}{0.5\hsize}
\includegraphics[height=6.5cm,width=6.5cm,clip]{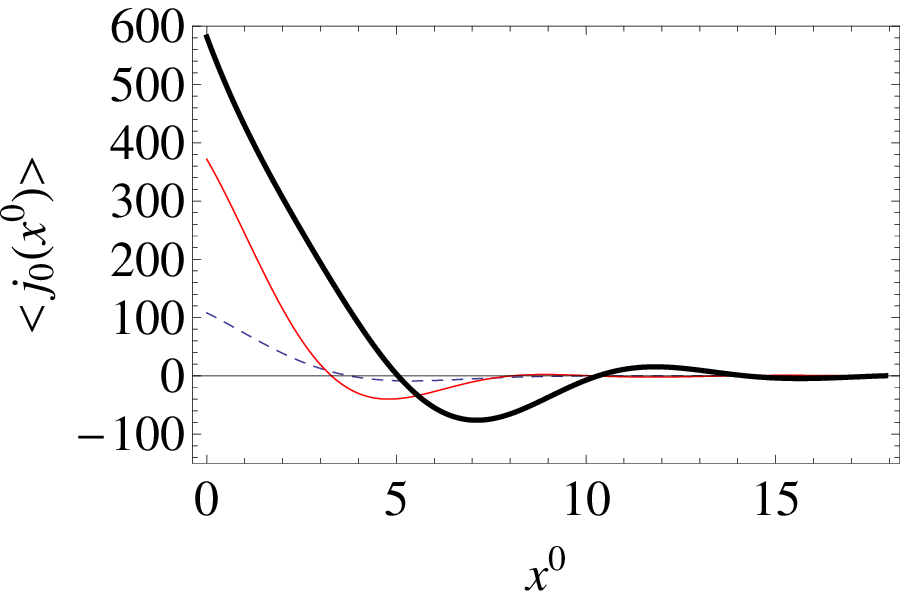}
\caption{\footnotesize{Dependence on chemical potential $\mu$ of time evolution of current density. The dashed line, the solid line, the thick solid line  show the case $\mu=5, 11$ and $20$ respectively.  ($B,H,T$)=($3,0.1,10$) for all the lines.}}
\label{fig8}
\end{minipage}
\end{tabular}
\end{center}
\end{figure}

In Fig. 7, we show the dependence on temperature T of 
the current density.
It depends on the temperature only through the initial
distribution function $h(k, \mu, T)$. As the temperature is higher,
the initial current density becomes larger.
We expect the oscillatory behavior will be more
pronounced for low temperature case and Fig. 7 shows the
tendency.
When the temperature $T$ is small compared with the mass scale
$m_\phi$, the oscillation period is determined by the inverse of mass 
difference $m_2-m_1$.  When temperature $T$ is larger than 
the mass scale,
the period will be proportional to $\frac{T}{m_\phi(m_2-m_1)}$.
Therefore, when the temperature $T$ is larger than $m_\phi$, 
the oscillation
period becomes large. 
We also show $\mu$ dependence of the current density in Fig. 8.
The chemical potential  
$\mu$ also influences the current density at 
the initial time. As the chemical potential becomes larger,
the initial current density becomes larger.

The effect of the large chemical potential 
on the time dependence of the current density 
is very different from that of the small chemical potential. 
In Fig. 9, we pay attention to the damping 
speed and observe the distinctive behavior 
between the two cases, i.e.,  $\mu > m_\phi$ and 
$m_\phi < \mu$. 
We compare the time dependence of the current density 
normalized by their initial values.
When the chemical potential exceeds the
mass scale $m_\phi$ (thin solid line)
the oscillatory behavior lasts much longer than the
case with the small chemical potential (thick solid line). 
The damping behavior $\sim e^{-3 H x^0}$ which is  expected from the simple
volume expansion of the universe is also shown with the dotted line.
The exponential damping rate of the current density for the large chemical 
potential case ($\mu = 20$) accords with the one expected from the volume expansion. 
The damping effect due to the destructive interference can not   
be seen when the chemical potential is greater than $m_\phi$. 
As shown in Fig.3, the momentum distribution has a pole at a certain momentum
satisfying the condition $\mu=\omega(k)$. 
Therefore the absolute value of the distribution function $h$ is very large 
within the small range of the momenta around the pole.  
From the behavior of the distribution function,
one concludes that the contribution only from a certain momentum region is 
dominant for the case $m_{\phi}< \mu$ and the oscillation period of the
current density is fixed even after integrating the distribution ($h$) $\times$
time evolution factor ($V$) over all the momenta. 
\begin{figure}
\begin{center}
\includegraphics[height=6.5cm,width=6.5cm,clip]{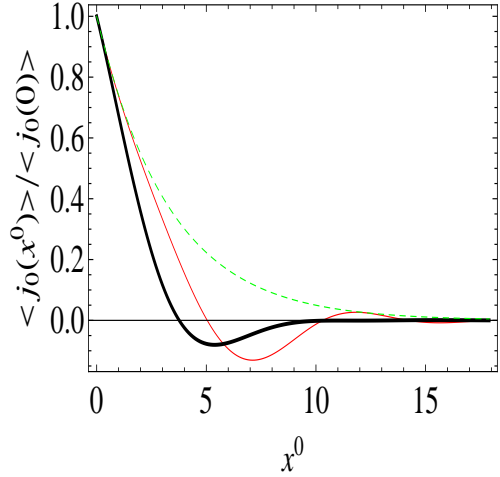}
\caption{We show the time dependence of the current densities normalized by their initial values.  The case with  $\mu > m_{\phi}$ and
the case with $\mu < m_\phi$ are shown.
The thin solid line shows the case with  
$\mu=20 $ and the thick solid line shows the case with
$\mu = 5$, respectively. We choose $m_{\phi}=10$ and 
$(B, H, T)=(3, 0.1, 10)$ for both cases.
For comparison, with dashed line, we show the time dependence for 
the inverse of the universe's volume, i.e.,
$e^{-3 H x^0}$ with $H=0.1$.}
\end{center}
\end{figure}
\section{Conclusion and discussion}
We have studied how the primordial matter and anti-matter asymmetry in expandinguniverse evolves under the influence of the particle number violating interactions. 
To investigate its time variation, we have introduced the complex scalar field with the well-defined particle number density.  The Lagrangian of the scalar field includes the mass term which breaks the particle number  conservation.
Such mass term can lead to the time evolution of the asymmetry existing at the beginning, however, it cannot produce the asymmetry itself.  
We have assumed that the particle number breaking mass term is turned on when the universe begins to expand. For the numerical calculation, 
we assume the universe
expands exponentially with respect to time.
Under the assumptions, the expectation value of the particle number density in later time is obtained and its formula has been given in an analytic form with the special function. 
The particle number density is written in terms of the momentum integration of the time dependent function $V(\eta, \eta_0, k)$ weighted with the distribution function (see Fig.1).
  By specifying the chemical potential and temperature in the density matrix, we have determined the initial condition for the particle number density. We have numerically calculated the evolution of the density and showed various cases 
with changing the expansion rate of the universe, the value of the particle number violating mass term, chemical potential and temperature. In particular, we have paid attention to the speed of decreasing of the particle number density.  When the particle number is conserved, the density decreases in inversely 
proportional to the volume of the universe.
 When the particle number violating interaction is turned on,
the behavior of the density is very different from that of the case without the interaction. 

There are two typical cases.  When the chemical potential is smaller than the mass of the complex scalar, besides the damping effect due to the expansion, the interference of contributions from various momenta also reduces the particle number density . 
On the other hand, when the chemical potential is larger than the scalar mass, the contribution from a certain momentum region is dominant.  The resulting particle number density oscillates with a definite frequency in addition to the damping.

The phenomena of the single frequency dominance is 
related to the fact that the distribution function for the complex scalar boson has a pole at some momentum. 
In contrast to the case with small chemical potential, the interference does not occur and the density continues to oscillate over the time until
the density itself is suppressed by the expansion of the universe.

The phenomena of the decoherence with the interference and the coherence with large chemical potential will have some 
impact on concrete scenarios of matter and anti-matter problem.
The decoherence effect has some impact on the dark matter problem.
In the scenario of the asymmetric dark matter, the dark matter is a remnant of the matter and anti-matter annihilation.  The matter and anti-matter oscillation, if it exists, can change the amount of the dark matter \cite{Cirelli:2011ac}, \cite{Tulin:2012re}.  
If the decoherence occurs and the primordial
asymmetry is washed out, the amount of the matter becomes nearly equal to the one of the anti-matter. 
The pair annihilation of matter and anti-matter leads to the further reduction 
of the dark matter.

Let us consider the case that the sign of the primordial particle number asymmetry is positive and is the same as that of the present asymmetry. 
Suppose the particle number violating mass term is so small that the sign of the asymmetry has been remained as positive. 
If this is the case, the strength of the particle number violating mass term will be determined with cosmological observation on the difference between the primordial asymmetry and the present one.
Although we have assumed in the numerical calculation that universe 
expands exponentially
with respect to time, it is also possible to extend to the case when the 
scale factor has more general dependence on time, such as a power law.
In principle, one can reduce the problem to solving the linear differential 
with the scale factor. The current density can 
be written in terms of the solutions and the formulae similar to Eq.(\ref{eq:current2})
will be obtained.  

  To extend the present model to a realistic one, we need to introduce new interactions and  new degrees of freedom so that the primordial 
density can be generated.
\appendix
\section{Approximate formulae for the small H limit}
In this appendix, we derive the approximate formulae of the particle number 
density, 
when the expansion rate $H$ is small in Eq.(\ref{eq:current2}). 
We start with the following integral representation for Hankel functions
\cite{Watson}.
\bea
\omega^R_{\lambda}(-x)&=&-\frac{1}{\pi} \int_R d \zeta 
e^{i x \sin \zeta + i \lambda \zeta}, 
\label{eq:int}
\eea 
where $\frac{k}{H}=x>0$ with $\lambda=i \rho=i \frac{m}{H}$. 
We first derive the asymptotic 
form for the Hankel functions in the small $H$ limit. When $H$ is small, both $x$ and $\lambda$
are large. To obtain the approximate form, we can write,
\bea
\lambda= i \rho=i \sigma x, \quad
\sigma=\frac{m}{k}.
\eea
Using the integral representation for the Hankel functions, we obtain the
approximate form for them in large $x$ limit.
In the integral representation,  
$R$ denotes the contour for the integration with respect to a complex
variable $\zeta=\xi+ i \eta$.
The contour is shown in Fig. 10.
\begin{figure}
\begin{center}
\includegraphics[width=8cm]{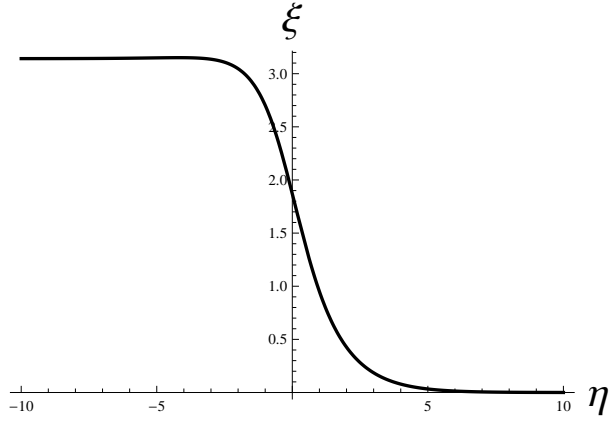}
\caption{The thick solid line shows the contour R.}
\end{center}
\label{fig:figcoun}
\end{figure}
The contour R is the curve which begins at $(\xi,\eta)=(\pi, -\infty)$ and ends at
$(0, \infty)$. On the contour R, $\xi$ varies within the range $[0,\pi]$.
One can rewrite Eq.(\ref{eq:int}) so that large $x$ limit is easily taken,  
\bea
\omega^R_{\lambda}(-x)&=&-\frac{1}{\pi} \int_R d \zeta e^{x f(\zeta)}, \nn \\
f(\zeta)&=&i \sin \zeta-\sigma \zeta=u(\xi, \eta) + i v(\xi, \eta),
\eea
where the real part $u$ and imaginary part $v$ of $f$ are given respectively
by,
\bea
u(\xi, \eta)&=&-\cos \xi \sinh \eta-\sigma \xi, \nn \\
v(\xi, \eta)&=& \sin \xi \cosh \eta-\sigma \eta.  
\eea 
We apply the steepest descent method and obtain 
the approximate form in the large $x$ limit.
We first find a saddle point of $u$. 
\bea
\frac{\partial u}{\partial \xi}=0 , \quad \frac{\partial u}{\partial \eta}=0.
\eea
The following conditions are satisfied at the saddle point.
\bea
\sin \xi \sinh \eta-\sigma=0, -\cos \xi \cosh \eta=0.
\eea
The saddle point for $u$ which lies in the range $\xi \in [0, \pi]$ is
\bea
\zeta_{0R}=(\xi_0, \eta_0)=(\frac{\pi}{2}, \sinh^{-1} \sigma).
\eea
Along the curve which passes the saddle point $\zeta_{0R}$ with steepest descent, $v$ is constant.
\bea
v(\xi,\eta)=v(\xi_0, \eta_0).
\label{eq:v}
\eea
One can solve Eq.(\ref{eq:v}) and obtain $\xi$ as a function of $\eta$. 
\bea
\xi_R(\eta)=\theta(\eta-\eta_0) \sin^{-1}(\frac{\cosh \eta_0-\sigma(\eta-\eta_0)}{\cosh \eta}) + 
\theta(\eta_0-\eta) [\pi-\sin^{-1}(
\frac{\cosh \eta_0-\sigma(\eta-\eta_0)}{\cosh \eta})]. \nn \\
\eea
The contour $\xi_R(\eta)$ is shown in Fig. 10 
as the thick solid curve. 
One can rewrite the contour integration of
the integral representation in Eq.(\ref{eq:int}) as,
\bea
\omega^R_{\lambda}(-x)=-\frac{1}{\pi} e^{x (u(\xi_0,\eta_0)+ iv(\xi_0,\eta_0))}
\int_{-\infty}^{+\infty} d \eta (\frac{d \xi_R }{d \eta}+i)
e^{x \{u(\xi_R(s),\eta)-u(\xi_0,\eta_0)\}}.
\eea
We carry out the integration with Gaussian approximation.
One expands the real part of $f(\zeta)$ around at the saddle point.
\bea
u_R(\eta)\equiv u(\xi_R(\eta), \eta)=u(\xi_0,\eta_0)+\frac{1}{2!} 
\frac{d^2 u_R}{d \eta^2}|_{\eta=\eta_0}(\eta-\eta_0)^2+...
\eea
Truncating the series up to the term quadratic with respect to $\eta-\eta_0$ and replacing $
\frac{d \xi_R}{d \eta}$ with 
$\frac{d \xi_R}{d \eta}\Bigr{|}_{\eta=\eta_0}$,
we obtain
\bea
\omega^R_{\lambda}(-x) &\sim& -\frac{1}{\pi} e^{x(u(\xi_0, \eta_0)+i v(\xi_0, \eta_0))}
(\frac{d \xi_R}{d \eta}+i)\Bigr{|}_{\eta=\eta_0}
\int_{-\infty}^{\infty} d\eta e^{\frac{x}{2} 
\frac{d^2 u_R}{d \eta^2}\Bigr{|}_{\eta=\eta_0}
(\eta-\eta_0)^2},\nn \\
&=& -\frac{1}{\pi} e^{x f(\zeta_0)} (\frac{d \xi_R}{d \eta}+i)\Bigr{|}_{\eta=\eta_0} 
\sqrt{\frac{2 \pi}{-x \frac{d^2 u_R}{d \eta^2}\Bigr{|}_{\eta=\eta_0}}}.
\eea
One finds, 
\bea
&&\frac{d \xi_R}{d \eta}\Bigr{|}_{\eta=\eta_0}=-1, \nn \\
&&\frac{d^2 u_R}{d \eta^2}\Bigr{|}_{\eta=\eta_0}=
-2 \cosh \eta_0. 
\eea
Therefore, for small $H$ limit, the Hankel function is given as,
\bea
\omega^R_{\lambda}(-x)&=& \sqrt{\frac{2}{ \pi x}}\left(\frac{1}{1+\sigma^2}\right)^{\frac{1}{4}}
e^{-i\frac{\pi}{4}-\frac{\sigma x \pi}{2}+ i x(\sqrt{1+\sigma^2}-\sigma \sinh^{-1} 
\sigma)}.
\eea
Since we derive the approximate form of the Hankel functions, one
can just substitute it into Eq.(\ref{eq:WZ}) and Eq.(\ref{eq:current2}).
We note the Eq.(\ref{eq:current2}) is independent of the normalization of the
solution. Therefore one can simply substitute 
\bea
H_{i \rho_n}(\eta) \simeq 
\left(\frac{1}{a^2}+ \sigma_n^2 \right)^{-\frac{1}{4}}e^{i x f(a,\sigma_n)},  
\eea
where $f(a, \sigma_n)$ is defined as,
\bea
f(a,\sigma_n)=\sqrt{\frac{1}{a^2}+\sigma_n^2}-\sigma_n
 \sinh^{-1} \sigma_n a.
\eea
One also obtains the derivative of $H_{ i \rho_n}$,
\bea
H_{i \rho_n}^\prime(\eta) \simeq i a
\left(\frac{1}{a^2}+ \sigma_n^2 \right)^{\frac{1}{4}}e^{i x f(a,\sigma_n)}.  
\eea
Using the results, one obtains the functions in Eq.(\ref{eq:WZ}).
\bea
W_n[\eta, \eta_0]&\simeq & 2 i
\left(\frac{1+\sigma_n^2}{\frac{1}{a^2}+\sigma_n^2} \right)^{\frac{1}{4}}
\cos x(f(a, \sigma_n)-f(1, \sigma_n)), \label{eq:WN} \\
Z_n[\eta, \eta_0]&\simeq & -2 i 
\left((1+\sigma_n^2)(\frac{1}{a^2}+\sigma_n^2) \right)^{-\frac{1}{4}}
\sin x(f(a, \sigma_n)-f(1, \sigma_n)),
 \label{eq:ZN}\\
W^\prime_n[\eta, \eta_0]&\simeq & 2 i a \left((1+\sigma_n^2)(\frac{1}{a^2}+\sigma_n^2) \right)^{\frac{1}{4}}
\sin x(f(a, \sigma_n)-f(1, \sigma_n)), \label{eq:WPN}\\
Z^\prime_n[\eta,\eta_0]& \simeq & 2 i a
\left(\frac{\frac{1}{a^2}+\sigma_n^2}{1+\sigma_n^2} \right)^{\frac{1}{4}}
\cos x(f(a, \sigma_n)-f(1, \sigma_n)). \label{eq:ZPN}
\eea
One substitutes the approximate formulas for the functions given in 
Eq.(\ref{eq:WN})-(\ref{eq:ZPN}) and
obtains,
\bea
&&<j_0(x^0)>= e^{-3 H x^0}\int \frac{d^3 k}{(2 \pi)^3}
\frac{\sinh \beta \mu}{ (\cosh \beta \omega(k)-\cosh \beta \mu)}\nn \\
&& \frac{1}{2}\Bigl{[}\Bigl{\{}\left(
\frac{\omega_1 \omega_1(x^0)}{\omega_2 \omega_2(x^0)}\right)^{\frac{1}{2}}+
\left(\frac{\omega_2 \omega_2(x^0)}{\omega_1 \omega_1(x^0)}
\right)^{\frac{1}{2}}
\Bigr{\}} 
\sin x(f(a, \sigma_1)-f(1, \sigma_1)) \sin x (f(a, \sigma_2)-f(1, \sigma_2))
 \nn\\
&+&\Bigl{\{}
\left(\frac{\omega_1 \omega_2(x^0)}{\omega_2 \omega_1(x^0)}\right)^{\frac{1}{2}}+
\left(\frac{\omega_2 \omega_1(x^0)}{\omega_1 \omega_2(x^0)}
\right)^{\frac{1}{2}}
\Bigr{\}} 
\cos x(f(a, \sigma_1)-f(1, \sigma_1)) \cos x (f(a, \sigma_2)-f(1, \sigma_2)) 
\Bigr{]}, \nn \\
\label{eq:asmallimit}
\eea 
where $\omega_i(x^0)$ $(i=1,2)$ are time dependent energies defined as,
\bea
\omega_i(x^0)=\sqrt{\frac{k^2}{a(x^0)^2}+m_i^2},
\eea
while $\omega_i$ is independent of the time.
\bea
\omega_i=\omega_i(x^0=0)=\sqrt{k^2+m_i^2}.
\eea
In the vanishing limit of $H$, one obtains,
\bea
x(f(a, \sigma_n)-f(1, \sigma_n)) \simeq -\omega_n x^0.
\eea
Therefore, in the limit, the current density is given as follows,
\bea
<j_0(x^0)>&=& \int \frac{d^3 k}{(2 \pi)^3}
\frac{\sinh \beta \mu}{ (\cosh \beta \omega(k)-\cosh \beta \mu)}\nn \\
&& \Bigl{[} \cos (\omega_1-\omega_2) x^0+\frac{1}{2}\Bigl{\{}
\frac{\omega_1}{\omega_2}+
\frac{\omega_2}{\omega_1}
-2\Bigr{\}} 
\sin \omega_1 x^0  \sin \omega_2 x^0
 \Bigr{]}. \nn \\
\label{eq:azerolimit}
\eea
\section{The formula for the case $B=0, \alpha_2 \rightarrow 0$}
In this appendix, we give the outline of 
the derivation for the vanishing limit
of the particle number violating mass term, i.e., $B \rightarrow 0$ and $\alpha_2 \rightarrow 0$.  
In this limit,
two mass eigen values of the real scalars are degenerate and 
one can set $\bar{m}_1(x^0)=\bar{m}_2(x^0)$ in Eq.(\ref{eq:fg}).
Then one can readily show the following equation.
\bea
\frac{d}{dx^0} \log \langle j(x^0) \rangle=-3 \frac{d}{d x^0} \log a(x^0),
\eea
which leads to Eq.(\ref{eq:bzerolimit}). 
%
\begin{acknowledgments}
T. M. was supported by KAKENHI, Grant-in-Aid for 
Scientific Research(C) No.22540283 from JSPS, Japan.
\end{acknowledgments}


\begin{thebibliography}{00}
\bibitem{Sakharov:1967dj} 
  A.~D.~Sakharov,
  Pisma Zh.\ Eksp.\ Teor.\ Fiz.\  {\bf 5}, 32 (1967)
  [JETP Lett.\  {\bf 5}, 24 (1967)]
  [Sov.\ Phys.\ Usp.\  {\bf 34}, 392 (1991)]
  [Usp.\ Fiz.\ Nauk {\bf 161}, 61 (1991)].
\bibitem{Yoshimura:1978ex} 
  M.~Yoshimura,
  Phys.\ Rev.\ Lett.\  {\bf 41}, 281 (1978)
  [Erratum-ibid.\  {\bf 42}, 746 (1979)].
\bibitem{Fukugita:1986hr} 
  M.~Fukugita and T.~Yanagida,
  Phys.\ Lett.\ B {\bf 174}, 45 (1986).
\bibitem{Hasegawa:2003vh} 
  K.~Hasegawa,
  Phys.\ Rev.\ D {\bf 69}, 013002 (2004)
  [hep-ph/0309098].
\bibitem{Dimopoulos:1978kv} 
  S.~Dimopoulos and L.~Susskind,
  Phys.\ Rev.\ D {\bf 18}, 4500 (1978).
\bibitem{Affleck:1984fy} 
  I.~Affleck and M.~Dine,
  Nucl.\ Phys.\ B {\bf 249}, 361 (1985).
\bibitem{Takeuchi:2010tm} 
  T.~Takeuchi, A.~Minamizaki and A.~Sugamoto,
  arXiv:1008.4515 [hep-ph].
\bibitem{Schwinger:1960qe} 
  J.~S.~Schwinger,
  J.\ Math.\ Phys.\  {\bf 2}, 407 (1961).
\bibitem{Bakshi:1962dv} 
  P.~M.~Bakshi and K.~T.~Mahanthappa,
  J.\ Math.\ Phys.\  {\bf 4}, 1 (1963).
\bibitem{Bakshi:1963bn} 
  P.~M.~Bakshi and K.~T.~Mahanthappa,
  J.\ Math.\ Phys.\  {\bf 4}, 12 (1963).
\bibitem{Keldysh:1964ud} 
  L.~V.~Keldysh,
  Zh.\ Eksp.\ Teor.\ Fiz.\  {\bf 47}, 1515 (1964)
  [Sov.\ Phys.\ JETP {\bf 20}, 1018 (1965)].


\bibitem{Ramsey:1997qc} 
  S.~A.~Ramsey and B.~L.~Hu,
  Phys.\ Rev.\ D {\bf 56}, 661 (1997)
  [gr-qc/9706001].
\bibitem{Calzetta:1986cq} 
  E.~Calzetta and B.~L.~Hu,
  Phys.\ Rev.\ D {\bf 37}, 2878 (1988).
\bibitem{CalzettaHu}
Nonequilibrium Quantum Field Theory, E. Calzetta and B.-L. Hu,
Cambridge University Press. (2008) 1-535.
\bibitem{Cirelli:2011ac} 
  M.~Cirelli, P.~Panci, G.~Servant and G.~Zaharijas,
  JCAP {\bf 1203}, 015 (2012)
  [arXiv:1110.3809 [hep-ph]].
\bibitem{Tulin:2012re} 
  S.~Tulin, H.~-B.~Yu and K.~M.~Zurek,
  JCAP {\bf 1205}, 013 (2012)
  [arXiv:1202.0283 [hep-ph]].
\bibitem{Watson} 
A treatise on the theory of Bessel functions, G. N. Watson,
Cambridge University Press. (1966) 1-804.
\end{thebibliography}
\end{document}